\documentclass[prd,amsmath,amssymb,superscriptaddress,preprintnumbers,twocolumn,showkeys,showpacs,10pt]{revtex4-1}

\pdfoutput=1
\usepackage{graphicx}
\usepackage{subfigure}
\usepackage{float}

\usepackage{dcolumn}
\usepackage{bm}
\usepackage{amssymb}
\usepackage{latexsym}
\usepackage{booktabs}
\usepackage{amsmath}
\usepackage{multirow}
\usepackage{url}
\usepackage{changes}
\usepackage{soul} 
\usepackage{color, xcolor}

\usepackage[colorlinks=true, linkcolor=red, citecolor=blue]{hyperref}

\usepackage[normalem]{ulem}
\usepackage{array}
\usepackage{enumerate}

\def\be{\begin{equation}}
\def\ee{\end{equation}}

\def\bea{\begin{eqnarray}}
\def\eea{\end{eqnarray}}
\newcommand{\bes}{\begin{equation*}}
\newcommand{\ees}{\end{equation*}}
\newcommand{\beqa}{\begin{eqnarray}}
\newcommand{\eeqa}{\end{eqnarray}}
\newcommand{\ksm}{~{\rm km}~{\rm s}^{-1}~{\rm Mpc}^{-1}}

\begin{document}

\title{Synergy between CSST galaxy survey and gravitational-wave observation: Inferring the Hubble constant from dark standard sirens}

\author{Ji-Yu Song}
\affiliation{Key Laboratory of Cosmology and Astrophysics (Liaoning) \& College of Sciences, Northeastern University, Shenyang 110819, China}
\author{Ling-Feng Wang}
\affiliation{Key Laboratory of Cosmology and Astrophysics (Liaoning) \& College of Sciences, Northeastern University, Shenyang 110819, China}
\author{Yichao Li}
\affiliation{Key Laboratory of Cosmology and Astrophysics (Liaoning) \& College of Sciences, Northeastern University, Shenyang 110819, China}
\author{Ze-Wei Zhao}
\affiliation{Key Laboratory of Cosmology and Astrophysics (Liaoning) \& College of Sciences, Northeastern University, Shenyang 110819, China}
\author{Jing-Fei Zhang}
\affiliation{Key Laboratory of Cosmology and Astrophysics (Liaoning) \& College of Sciences, Northeastern University, Shenyang 110819, China}
\author{Wen Zhao}
\affiliation{CAS Key Laboratory for Researches in Galaxies and Cosmology, Department of Astronomy, University of Science and Technology of China, Chinese Academy of Sciences, Hefei, Anhui 230026, China}
\affiliation{School of Astronomy and Space Sciences, University of Science and Technology of China, Hefei 230026, China}
\author{Xin Zhang}\thanks{Corresponding author.\\zhangxin@mail.neu.edu.cn}
\affiliation{Key Laboratory of Cosmology and Astrophysics (Liaoning) \& College of Sciences, Northeastern University, Shenyang 110819, China}
\affiliation{National Frontiers Science Center for Industrial Intelligence and Systems Optimization, Northeastern University, Shenyang 110819, China}
\affiliation{Key Laboratory of Data Analytics and Optimization for Smart Industry (Ministry of Education), Northeastern University, Shenyang 110819, China}

\begin{abstract}

Gravitational waves (GWs) from compact binary coalescences encode the absolute luminosity distances of GW sources. Once the redshifts of GW sources are known, one can use the distance-redshift relation to constrain cosmological parameters. One way to obtain the redshifts is to localize GW sources by GW observations and then use galaxy catalogs to determine redshifts from a statistical analysis of redshift information of the potential host galaxies, commonly referred to as the dark siren method. The third-generation (3G) GW detectors are planned to work in the 2030s and will observe numerous compact binary coalescences. Using these GW events as dark sirens requires high-quality galaxy catalogs from future sky survey projects. The China Space Station Telescope (CSST) will be launched in 2024 and will observe billions of galaxies within a 17500 deg$^2$ survey area with redshift up to $z\sim 4$, providing photometric and spectroscopic galaxy catalogs. In this work, we simulate the CSST galaxy catalogs and the 5-year GW data from the 3G GW detectors and combine them to infer the Hubble constant ($H_0$). Our results show that the measurement precision of $H_0$ could reach the sub-percent level, meeting the standard of precision cosmology. We conclude that the synergy between CSST and the 3G GW detectors is of great significance in measuring the Hubble constant.


\end{abstract}

\pacs{98.80.Es, 95.36.+x, 98.80.-k, 04.80.Nn, 98.58.Bz}
\keywords{CSST galaxy survey, galaxy catalogs, gravitational waves, dark sirens, the Hubble constant}

\maketitle

\section{Introduction}

In the past two decades, sky survey projects, such as the Sloan Digital Sky Survey (SDSS) \cite{SDSS:2000hjo}, have achieved important results and initiated a new era of exploring fundamental physics through astronomical observations. To make further improvements, several next-generation ground-based and space-borne telescopes, such as the Large Synoptic Survey Telescope (LSST) \cite{LSSTScience:2009jmu,LSST:2008ijt}, the {Euclid} space mission \cite{EUCLID:2011zbd}, the Wield Field Infrared Survey Telescope (WFIRST) \cite{WFIRST-web}, and the China Space Station Telescope (also known as the Chinese Survey Space Telescope, CSST) \cite{Cao:2017ph,Cao:2021bqm, Cao:2021ykj}, have been scheduled to implement.

CSST is a space telescope with a two-meter aperture, which has been planned to be launched in 2024 and will enter the same orbit as the China Manned Space Station. The Chinese Space Station Optical Survey (CSS-OS) is the major science project operated by CSST \cite{Cao:2017ph}, and it will simultaneously perform both the photometric redshift (photo-z) and slitless spectroscopic redshift (spec-z) surveys, covering a total sky area of 17500 deg$^2$ in about ten years with a view field of 1.1 deg$^2$.

Several papers have forecasted what role CSST will play in the future cosmological research \cite{Gong:2019yxt,Cao:2021bqm,Cao:2021ykj,Zhou:2021ojw,Chen:2022wll,Lin:2022aro,Xu:2022gzj,Deng:2022mmu,Zhou:2022vjr,Wang:2022uel,Miao:2022hyp,Li:2022bfc,Liu:2022mur}. For example, Gong \emph{et al.} \cite{Gong:2019yxt} found that CSS-OS can improve the constraint precisions of the cosmological parameters by several times, compared with the current weak lensing and galaxy clustering surveys. Chen \emph{et al.} \cite{Chen:2022wll} found that the CSST galaxy clustering spectroscopic survey can powerfully constrain the Brans-Dicke (BD) theory and other modified gravity theories.
Lin \emph{et al.} \cite{Lin:2022aro} used the mock data from the CSST photometric galaxy clustering and cosmic shear surveys to constrain the total neutrino mass and obtained a comparable result to the \emph{Planck} result if the baryonic effect is ignored, much better than the results from current photometric surveys. Li \emph{et al.} \cite{Li:2022bfc} forecasted that the CSST ultra-deep field observation can detect $\sim$ 1800 type Ia supernovae (SNe Ia) at $z<1.3$, and the SNe Ia samples of CSST could significantly improve the constraints on cosmological parameters, compared with the Pantheon sample.

In addition to the above aspects, we note that the CSS-OS galaxy catalog can provide the redshift information for the well-localized gravitational-wave (GW) events whose luminosity distances ($d_{\rm L}$) encoded in the GW waveforms (known as standard sirens \cite{Schutz:1986gp,Holz:2005df}), and thus has potential to provide measurements for cosmological parameters via the distance-redshift ($d_{\rm L}$-$z$) relation. In this paper, we wish to study what role the synergy between the CSS-OS and future GW observations will play in measuring cosmological parameters.

The typical GW events used for standard sirens are compact binary coalescences (CBCs), such as binary neutron star (BNS) coalescences and binary black hole (BBH) coalescences.
The redshifts of CBCs can be obtained mainly in two ways \cite{Schutz:1986gp}. One way is to observe the electromagnetic (EM) signals (EM counterparts) emitted when binaries merge, and these standard sirens are known as bright sirens \cite{Nissanke:2009kt,Tamanini:2016uin,LIGOScientific:2017adf,Cai:2017yww,DiValentino:2017clw,Zhao:2018gwk,DiValentino:2018jbh,Yang:2019vni,DES:2019ccw,Wang:2019tto,Zhao:2019gyk,Chen:2020zoq,Qi:2021iic,Chen:2020gek,Wang:2021srv,Jin:2022tdf,Jin:2023zhi,Jin:2023tou,Han:2023exn}. The other way is to use galaxy catalogs to provide redshifts of the potential host galaxies within the localization regions of the GW events and these standard sirens are known as dark sirens \cite{Nishizawa:2012vk,Cai:2017buj,Chen:2020dyt,LIGOScientific:2019zcs,Zhu:2021bpp,Jin:2023sfc,Jin:2022qnj}. 
Limited by the EM-counterpart observations, only a small fraction of GW events can be used as bright sirens.

Until now, the only available bright siren (GW170817) provides a $\sim$ $14\%$ measurement for $H_0$ \cite{LIGOScientific:2017vwq}. 47 CBCs from the third Gravitational-Wave Transient Catalog (GWTC-3), together with the {\tt GLADE+} galaxy catalog \cite{Dalya:2018cnd,Dalya:2021ewn}, 
are used as dark sirens and provide a $\sim19\%$ measurement for $H_0$ \cite{LIGOScientific:2021aug}.

In the 2030s, the third-generation (3G) GW detectors, i.e., the Einstein Telescope (ET) \cite{Punturo:2010zz} and the Cosmic Explorer (CE) \cite{LIGOScientific:2016wof}, are planned to work. Previous works show that the 3G GW detectors could detect $\mathcal{O}(10^6)$ BNS coalescences in the 10-year observation, of which only $\mathcal{O}(10^3)$ BNS coalescences could be used as bright sirens \cite{Wang:2018lun,Zhang:2018byx,Zhang:2019loq,Zhang:2019ylr,Zhang:2019ple,Li:2019ajo,Jin:2020hmc,Jin:2021pcv,Hou:2022rvk,Wu:2022dgy}. Most of the BNS coalescences and all the stellar-mass BBH coalescences have no EM counterparts and cannot serve as bright sirens. Therefore, an important question is how these numerous CBCs could be used as dark sirens in cosmological parameter estimations. 

Compared with the second-generation GW detectors, the 3G GW detectors have the sensitivities improved by over one order of magnitude and could detect CBCs even up to $z$ $\sim$ 100 \cite{Evans:2021gyd}. Nevertheless, the existing galaxy catalog cannot meet the requirements of future dark-siren observations. The {\tt GLADE+} galaxy catalog used for the current dark sirens can be considered complete only at $z<0.011$ and the completeness falls to 20\% at $z\sim0.167$, limiting the capability of the dark-siren method in the era of 3G GW detectors. To realize the full potential of the dark siren method \cite{Dalya:2021ewn}, we need the next-generation sky survey projects to work with the 3G GW detectors.

CSST, a next-generation space telescope, is scheduled to complete its survey project CSS-OS in around 2034, presenting a comprehensive galaxy catalog for the 3G GW detectors. Compared with the {\tt GLADE+} galaxy catalog utilized in the GWTC-3 analysis, the CSS-OS galaxy catalog has notable advantages, such as extended magnitude limits and reduced redshift uncertainties. These enhancements will contribute to a more exhaustive and precise galaxy catalog, improving the ability to localize the host galaxies of dark sirens. In addition, CSST outperforms other Stage IV survey telescopes like LSST, {Euclid}, and WFIRST in some aspects \cite{Gong:2019yxt, article}. Therefore, it becomes imperative to investigate the synergy between CSST and the 3G GW detectors in precisely measuring cosmological parameters via the dark siren method.

In this paper, we only focus on measuring the Hubble constant ($H_0$) via dark sirens due to the following two reasons.
(i) $H_0$ as the first cosmological parameter has been measured for about one century. However, currently, the constraint on $H_0$ from the \emph{Planck} cosmic microwave background (CMB) data (assuming a $\Lambda$CDM model) and the direct measurement of $H_0$ using the cosmic distance ladder are in more than 4$\sigma$ tension, known as the $H_0$ tension \cite{Zhao:2017urm,Zhang:2017epd,Guo:2017qjt,Yang:2018euj,Guo:2018uic,Guo:2018ans,DiValentino:2019jae,DiValentino:2019ffd,Liu:2019awo,Zhang:2019cww,Ding:2019mmw,Feng:2019mym,Guo:2019dui,Xu:2020uws,Li:2020tds,Gao:2021xnk,Cai:2021wgv,Vagnozzi:2021gjh,Cao:2021zpf,Wang:2021kxc,Vagnozzi:2021tjv,Kamionkowski:2022pkx,Guo:2022,Wu:2022jkf,Zhang:2023gye,Dai:2023,Chen:2023}. The issue of how to precisely measure $H_0$ becomes one of the key questions in cosmology.
(ii) The main advantage of dark sirens in our work is to precisely measure the Hubble constant because we use the dark sirens at low redshifts in which the $d_{\rm L}$-$z$ relation is strongly sensitive to $H_0$ but insensitive to other cosmological parameters. 

This paper is organized as follows. Sec.~\ref{sec:catalog} introduces the method of simulating the CSS-OS galaxy catalog. Sec.~\ref{sec:SIMULATION OF THE GW EVENT CATALOG} introduces the method of simulating the GW events. In Sec.~\ref{sec:FIM}, we estimate the measurement errors of GW source parameters. Sec.~\ref{sec:Search} introduces the method of identifying the host galaxies of GW events. Sec.~\ref{sec:baye} introduces the Bayesian method used to infer $H_0$. In Sec.~\ref{sec:result}, we show the constraint results of $H_0$ and make detailed discussions. The conclusion is given in Sec.~\ref{sec:summary}. 

\section{Simulations of the CSS-OS galaxy catalog}\label{sec:catalog}

\subsection{Distance-redshift relation}
The luminosity distance ($d_{\rm L}$) of a source at redshift $z$ is given by
\begin{equation}\label{con:distance-redshift relation}
    \begin{aligned}
        d_{\rm L}(z)=c(1+z)\int^z_0 \frac{{\rm d}z'}{H(z')},
    \end{aligned}
\end{equation}
where $c$ is the speed of light and $H(z)$ is the Hubble parameter describing the expansion rate of the universe at redshift $z$. In this work, we consider the $\Lambda$CDM model as the fiducial model, and the form of $H(z)$ is given by
\begin{equation}\label{con:Hubble parameter}
    \begin{aligned}
        H(z)=H_0\sqrt{\Omega_{\rm m}(1+z)^3+1-\Omega_{\rm m}},
    \end{aligned}
\end{equation}
where $\Omega_{\rm m}$ is the current matter density parameter. The fiducial values of cosmological parameters are chosen as the \emph{Planck} 2018 TT,TE,EE+lowE results \cite{Planck:2018nkj} with $\Omega_{\rm m}=0.3166$ and $H_0=67.27\ {\rm km\ s^{-1}\ Mpc^{-1}}$.

\subsection{Completeness and sky coverage of the CSS-OS galaxy catalogs}\label{subsec:completeness}

In this work, we consider both the spec-z and photo-z surveys of CSST. 
The wavelength coverage of CSST's photo-z and spec-z survey are both a full range of 255--1000 nm, from near-ultraviolet to near-infrared.  
The CSST's photo-z survey has seven broad-band filters, i.e., $NUV$, $u$, $g$, $r$, $i$, $z$, and $y$, and the CSST's spec-z survey has three bands, i.e., $GU$, $GV$, and $GI$.

To mock the photo-z catalog of CSS-OS, we first need to know the number density and the luminosity distribution of the galaxies in the universe.
In this paper, we assume that the luminosity distribution of galaxies follows a Schechter function \cite{Schechter:1976iz},
\begin{equation}\label{con:Schechter function}
    \begin{aligned}
        {\rm d}n(L)=\phi^{*}(L/L^{*})^{\alpha}\exp(-L/L^{*}){\rm d}L/L^{*},
    \end{aligned}
\end{equation}
with $\phi^{*}=1.6\times10^{-2}\ h^3{\rm Mpc}^{-3}$ and $\alpha=-1.07$. $L$ is the galaxy luminosity and $L^{*}=1.2\times10^{10}\ h^{-2}L_{\odot}$ is a characteristic galaxy luminosity, with $L_{\odot}$ being the solar luminosity, corresponding to a solar absolute magnitude ($Mag_{\odot}=5.48$). ${\rm d}n(L)$ is the number density of galaxies within the luminosity interval $[L,L+{\rm d}L]$. The parameters of the Schechter function we adopt here are from the $B$-band measurements of nearby galaxies \cite{Gehrels:2015uga}. Since the wavelength of the $B$ band is close to that of CSS-OS's $g$ band, we use this Schechter function to describe the luminosity distribution of the galaxies in the $g$ band.

We integrate Eq.~(\ref{con:Schechter function}) to determine the galaxy number density $n_{\rm tot}$ in the universe,
\begin{equation}
    \begin{aligned}
        n_{\rm tot}=\int_{L_{\rm lower}}^{\infty}\phi^{*}(L/L^{*})^{\alpha}\exp(-L/L^{*}){\rm d}L/L^{*},
    \end{aligned}
\end{equation}
where $L_{\rm lower}$ is the lower luminosity cut-off for the dimmest galaxies in the universe. Following Ref.~\cite{Wang:2020dkc}, we assume $n_{\rm tot}\sim0.02\ {\rm Mpc^{-3}}$ in the comoving volume, and hence $L_{\rm lower}\sim 0.0125L^{*}$. 

After simulating the luminosities of galaxies in the universe, we could determine which galaxies could be observed by CSST's photo-z survey. We define an apparent magnitude threshold ($mag_{\rm th}$), and the galaxies whose apparent magnitudes lower than this threshold could be observed. For the $g$ band of the CSS-OS photo-z catalog, we have $mag_{\rm th}\sim 25.5$ \cite{2023MNRAS.523..876Q}. We convert the luminosities of galaxies to apparent magnitudes with the following transfer function,
\begin{equation}\label{con:magnitude function}
    \begin{aligned}
        mag(L,d_{\rm L},z, C)=&Mag_{\odot}-2.5{\rm log}_{10}^{(L/L_{\odot})}+5{\rm log}_{10}^{(d_{\rm L}/{\rm pc})}\\&-5+K(z,C),
    \end{aligned}
\end{equation}
where $mag$ is the apparent magnitude, and $K$ represents the $K$ correction. We employ the polynomial fitting functions from Ref.~\cite{Chilingarian:2010sy} to calculate the $K$ correction. CSS-OS has similar filters with the corresponding filters of SDSS and LSST, especially for the $u$, $g$, $r$, and $i$ bands \cite{Cao:2017ph}. Hence, we choose the $K$-correction function for SDSS's $g$ band to calculate the $K$ correction for CSST's $g$ band. Here $C$ in the $K$ correction represents the color value. Different color values correspond to different types of galaxies, with active star-forming galaxies (SFG) corresponding to $C<0.15$ mag, late-type spirals (S-late) corresponding to $0.4~{\rm mag}<C<0.6~{\rm mag}$, early-type spirals (S-early) corresponding to $0.58~{\rm mag}<C<0.70~{\rm mag}$, and luminous red galaxies (LRG) corresponding to $0.73~{\rm mag}<C<0.81~{\rm mag}$.

\begin{figure}
    \centering
    \includegraphics[width=8.6cm]{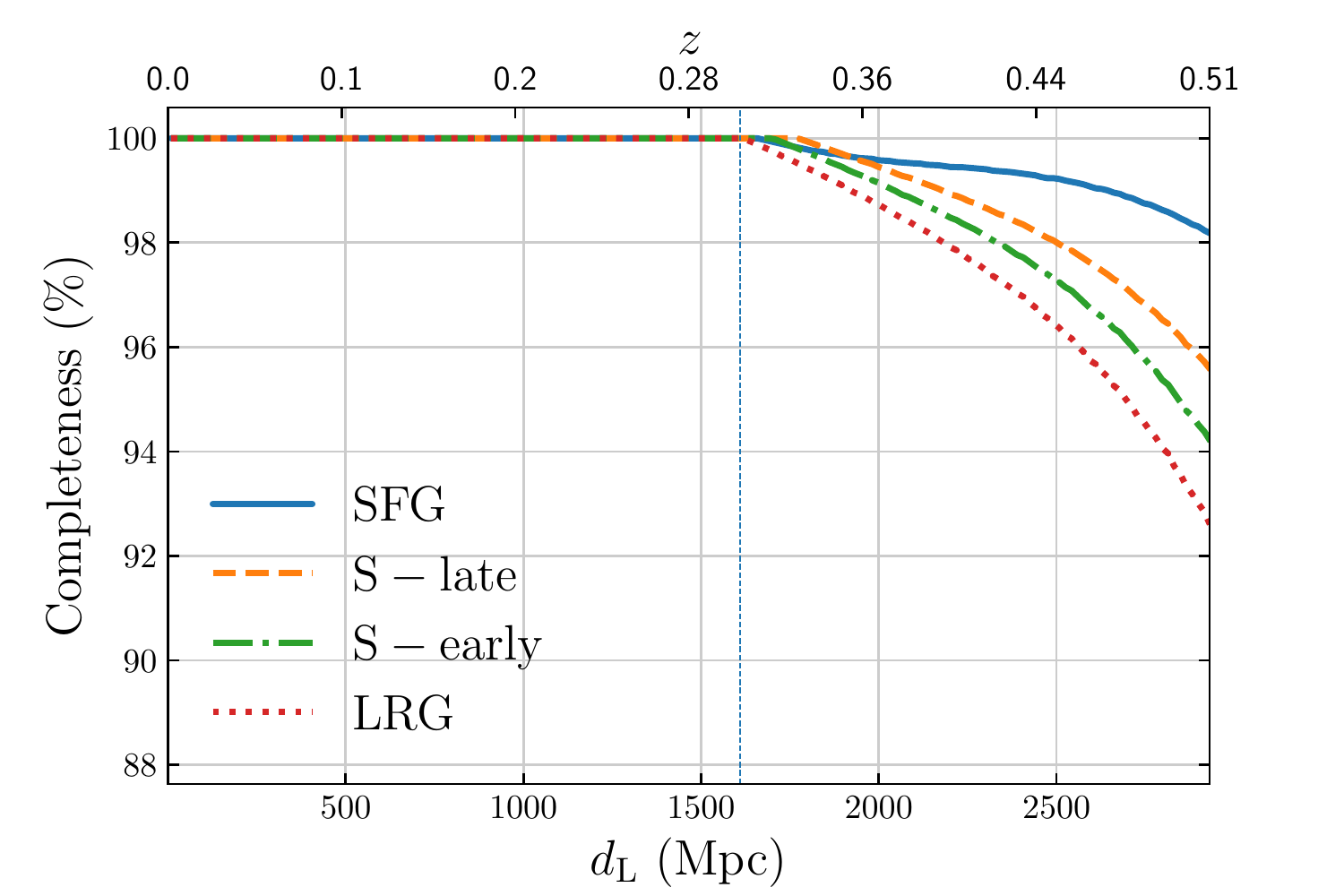}
    \caption{Completeness distribution of the CSS-OS photo-z galaxy catalog. Different lines represent different morphological types, shown as in legends. The blue vertical dashed line denotes $z = 0.3$.}
    \label{fig:completeness}
\end{figure}

Lastly, we can calculate the completeness distribution of the CSS-OS photo-z galaxy catalog. We divided $d_{\rm L}$ into different bins with a width of 17 Mpc, and the completeness of each $d_{\rm L}$ bin is defined as the fraction of the galaxies whose apparent magnitudes are lower than the apparent magnitude threshold $mag_{\rm th}$ \cite{LIGOScientific:2021aug}. In Fig.~\ref{fig:completeness}, we show the completeness distributions of the CSS-OS photo-z catalogs. We can clearly see that the completeness of the CSS-OS photo-z galaxy catalog is approximately 100\% at redshift 0.3. Therefore, we make the assumption that the photo-z CSS-OS galaxy catalog is complete at $z\leq 0.3$. As shown in Ref.~\cite{Gong:2019yxt}, the galaxy number densities inferred by the spec-z and photo-z measurements are consistent at $z\leq 0.1$. Thus, we consider the spec-z catalog to be complete at $z\leq 0.1$. In the subsequent analysis, we consider the photo-z survey within the redshift range $0.1<z\leq 0.3$ and consider the spec-z survey within the redshift range $z\leq0.1$.

For the sky coverage of CSS-OS, following Ref.~\cite{Yao:2023prr}, we remove the regions within $\pm 19.2~\deg$ of the galactic latitude and the ecliptic latitude and assume the remaining to be the coverage area of CSS-OS, which is about $17572~\deg^2$.

\subsection{Redshift uncertainties of galaxies}\label{subsec:redshift uncertainty}

According to Ref.~\cite{Cao:2017ph}, in the CSS-OS photo-z catalog, approximately 95\% of galaxies have a redshift fitting value with a deviation less than 0.05, while approximately 58\% of galaxies have a redshift fitting value with a deviation less than 0.02.
For simplicity, we consider the redshift instrumental uncertainties of the galaxies in the CSS-OS photo-z catalog with two extreme cases, i.e., the ``optimistic'' case with $\sigma_z(z)=0.02(1+z)$ and the ``conservative'' case with $\sigma_z(z)=0.05(1+z)$. For the CSS-OS spec-z catalog, we follow Ref.~\cite{Gong:2019yxt} and adopt $\sigma_z(z)=0.002(1+z)$.

\section{SIMULATION OF THE GW EVENT CATALOG}\label{sec:SIMULATION OF THE GW EVENT CATALOG}

\subsection{Masses}
For the distribution of the primary black hole's mass ($m_1$) in a BBH, we adopt the power law $+$ peak model \cite{LIGOScientific:2020kqk,KAGRA:2021duu}, given by
\begin{equation}
    \begin{aligned}
        p\left(m_{1}\right) =\left[\left(1-\lambda_{\text {peak }}\right) \mathcal{B}\left(m_{1}\right)+\lambda_{\text {peak }} \mathcal{G}\left(m_{1}\right)\right] \mathcal{S}\left(m_{1}\right),
    \end{aligned}
\end{equation}
with $\lambda_{\rm peak}=0.038$. $\mathcal{B}\left(m_{1}\right)\propto m_1^{-\alpha}$ is a normalized power-law distribution with the spectral index $\alpha=3.5$ and the high-mass cut-off $m_{\rm max}=44\ M_{\odot}$. $\mathcal{G}\left(m_{1}\right)$ is a Gaussian distribution with the mean $\mu_m=34\ M_{\odot}$ and the 1-$\sigma$ width $\sigma_m=5.69\ M_{\odot}$. $\mathcal{S}\left(m_{1}\right)$ is a smoothing function, given by
\begin{equation}
    \begin{aligned}
        \begin{array}{l}
\mathcal{S}(m_1)= \\
\left\{\begin{array}{ll}
0 & \text { if } m_1<m_{\min }, \\
\left[f\left(m_1-m_{\min }\right)+1\right]^{-1} & \text { if } m_{\min } \leq m_1<m_{\min }+\delta_{m}, \\
1 & \text { if } m_1 \geq m_{\min }+\delta_{m},
\end{array}\right.
\end{array}
    \end{aligned}
\end{equation}
with
\begin{equation}
    \begin{aligned}
        f(m)=\exp \left(\frac{\delta_{m}}{m}+\frac{\delta_{m}}{m-\delta_{m}}\right),
    \end{aligned}
\end{equation}
$m_{\rm min}=5\ M_{\odot}$, and $\delta_m=4.9\ M_{\odot}$.

The mass of the secondary black hole in BBH, denoted as $m_2$, is determined by $m_2=m_1 q$, where $q$ represents the mass ratio. The probability distribution of the mass ratio can be expressed as
\begin{equation}
    p(q) \propto q^{\beta} \mathcal{S}\left(q m_{1}\right),
\end{equation}
with $\beta=1.1$.

\subsection{Redshifts}\label{sec:Redshift}

The BBH merger rate (per comoving volume per year) in the observer frame can be expressed as \cite{Belgacem:2019tbw,Vitale:2018yhm,Yang:2021qge}
\begin{equation}
R_{\rm obs}(z)= \frac{R_{\rm m}(z)}{1+z},
\label{con:pz}
\end{equation}
where $R_{\rm m}(z)$ represents the merger rate in the source frame and is related to the binary system's formation rate by the time delay distribution,
\begin{equation}
    \begin{aligned}
    R_{\rm m}(z_{\rm m})=\int_{z_{\rm m}}^{\infty}dz_{\rm f}\frac{dt_{\rm f}}{dz_{\rm f}}R_{\rm f}(z_{\rm f})P(t_{\rm d}),
    \end{aligned}
\end{equation}
where $t_{\rm f}$ is the time when the binaries form, and is related to the redshift $z_{\rm f}$; $t_{\rm d}=t_{\rm f}-t_{\rm m}$ is the time delay in which $t_{\rm m}$ is the merger time and also the look-back time of the redshift $z_{\rm m}$; $P(t_{\rm d})$ is the distribution of the time delay, and we adopt the exponential form as \cite{Vitale:2018yhm}
 \begin{equation}
     P\left(t_{\rm d}\right)=\frac{1}{\tau} \exp \left(-t_{\rm d} / \tau\right),
 \end{equation}
with $\tau=100\ {\rm Myr}$. $R_{\rm f}(z)$ is the formation rate of the binary system, assumed to be proportional to the Madau-Dickinson (MD) star formation rate \cite{Madau:2014bja},
\begin{equation}
    \begin{aligned}
    R_{\rm f}(z)= A\frac{(1+z)^{2.7}}{1+[(1+z)/2.9]^{5.6}}.
    \end{aligned}
\label{equa:Rf}
\end{equation}
The coefficient $A$ is the normalization factor, determined by the BBH merger rate at $z=0$. For the BBH merger rate at $z=0$, we set $R_{\rm obs}(z=0)=R_{\rm m}(z=0)=23.9$ $\rm{Gpc}^{-3}$ $\rm{yr}^{-1}$ \cite{KAGRA:2021duu}. Integrating Eq.~(\ref{con:pz}) gives the number of BBH coalescence events per year,
\begin{equation}
    \begin{aligned}
        N_{\rm GW}=\int_{z_{\rm lower}}^{z_{\rm upper}} R_{\rm obs}(z)\frac{{\rm d}V_{\rm c}}{{\rm d}z}{\rm d}z,
    \end{aligned}
\end{equation}
where $z_{\rm lower}$ and $z_{\rm upper}$ represent the lower and upper limits of the redshift range considered in the analysis. $V_{\rm c}$ is the comoving volume.

According to our calculations, within the range of $z<0.3$ and the 17500 deg$^2$ sky coverage of CSS-OS, there are about 120 BBH coalescence events per year.

\subsection{Other GW source parameters}

The inclination angle $\iota$, the polarization angle $\psi$, and the coalescence phase $\psi_{\rm c}$ are randomly chosen in the range of ${\rm cos(\iota)}\in[-1,1]$, $\psi\in[0,2\pi]$, and $\psi_{\rm c}\in[0,2\pi]$. The angular locations of galaxies are typically labeled by declination (DEC) and right ascension (RA). In our work, we use the colatitude $\theta=\pi/2-{\rm DEC}$ and the longitude $\phi={\rm RA}$. For galaxies in the mock CSS-OS galaxy catalog, $\theta$ and $\phi$ are chosen in the coverage of CSS-OS, according to the distribution shown as follows,
\begin{equation}
    \begin{aligned}
        &p(\theta)\rm{d}\theta\propto\sin{\theta}\rm{d}\theta,\\
        &p(\phi)\rm{d}\phi\propto\rm{d}\phi.
    \end{aligned}
\end{equation}
Without loss of generality, we set the coalescence time to $t_{\rm c}=0$ \cite{Zhao:2017cbb}.

\section{GW detection and Fisher information matrix}\label{sec:FIM}

In this section, we calculate the signal-to-noise ratios (SNRs) of GW sources by simulating the mock GW signals and introduce the Fisher information matrix (FIM). For the GW detectors, we assume that ET is located in Europe and has three 10 km arms, forming a triangular configuration; there are two L-shaped CEs, one in the United States (CE1) with two 40 km arms and the other one in Australia (CE2) with two 20 km arms. We obtain the detailed locations of the GW detectors from Refs.~\cite{LIGOScientific:2018mvr,Ashton:2018jfp,Borhanian:2020ypi,2021SeiRL..92..352D}. In our analysis, we consider three scenarios, i.e., ET, CE1, and the detection network comprised of ET, CE1, and CE2, referred to as ET2CE.

The strain $h(t)$ of GW interferometers in the transverse-traceless gauge can be described by two independent polarization amplitudes,
\begin{equation}
h(t)=F_+(\theta, \phi, \psi)h_+(t)+F_\times(\theta, \phi, \psi)h_\times(t),
\end{equation}
where $F_{+}$ and $F_{\times}$ are antenna response functions, and their specific forms for ET and CEs are taken from Refs.~\cite{Zhao:2010sz,Jin:2020hmc}. 

Here, we adopt the GW waveforms in the inspiral phase of non-spinning BBH systems. Following Refs.~\cite{Blanchet:2004bb,Sathyaprakash:2009xs}, we use the restricted post-Newtonian (PN) approximation and calculate the waveform to the 3.5 PN order, obtaining the Fourier transform $\tilde{h}(f)$ of the time domain waveform $h(t)$, shown as followed,
\begin{equation}
\tilde{h}(f)=\mathcal{A}f^{-7/6}\exp[{\rm i}(2\pi ft_{\rm c}-\pi/4-2\psi_{\rm c}+2{\Psi}(f/2)-\varphi_{(2.0)})].
\label{equa:hf}
\end{equation}
The Fourier amplitude $\mathcal{A}$ is given by
\begin{align}
\mathcal{A}=&~~\frac{1}{d_{\rm L}}\sqrt{F_+^2\big(1+\cos^2(\iota)\big)^2+4F_\times^2\cos^2(\iota)}\nonumber \times\\
            &~~ \sqrt{5\pi/96}\pi^{-7/6}\mathcal{M}_{\rm c}^{5/6},
\label{equa:A}
\end{align}
where $d_{\rm L}$ is the luminosity distance of the GW event. $\mathcal{M}_{\rm c}=(1+z)M \eta^{3/5}$ is the observed chirp mass and $M=m_1+m_2$ is the total mass of the BBH. $\eta=m_1 m_2/M^2$ is the symmetric mass ratio. 
$\varphi_{(2.0)}$ and $\Psi(f)$ are given by \cite{Blanchet:2004bb,Sathyaprakash:2009xs}
\begin{equation}
    \varphi_{(2,0)}=\tan ^{-1}\left(-\frac{2 \cos (\iota) F_{\times}}{\left(1+\cos ^{2}(\iota)\right) F_{+}}\right),
\end{equation}
\begin{equation}
    \Psi(f)=-\psi_{\mathrm{c}}+\frac{3}{256 \eta} \sum_{i=0}^{7} \psi_{i}(2 \pi M f)^{(i-5) / 3},
\end{equation}
where the coefficients $\psi_i$ are detailed in Ref.~\cite{Sathyaprakash:2009xs}.

\begin{figure}
    \centering
    \includegraphics[width=8.6cm]{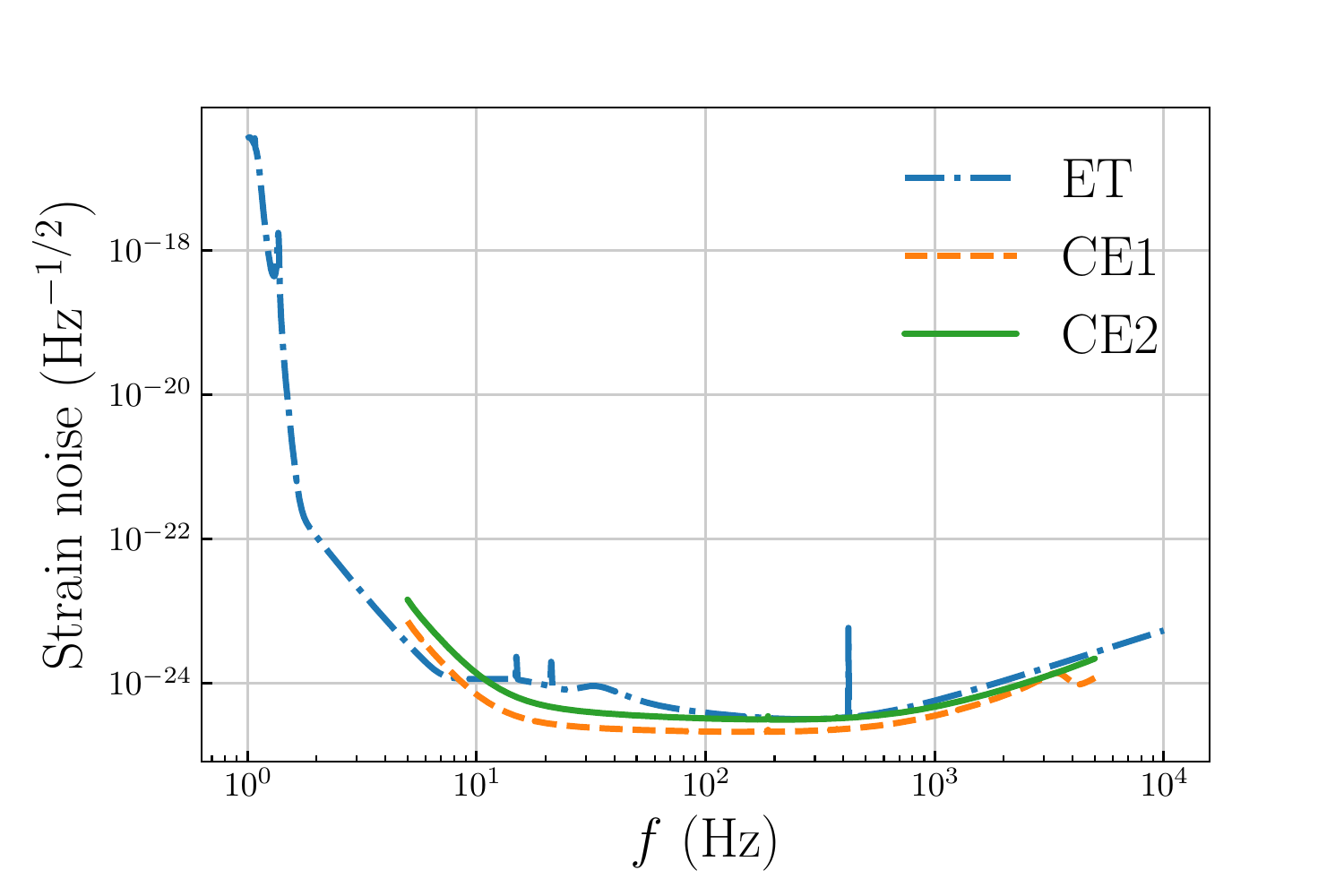}
    \caption{Sensitivity curves of the 3G GW detectors.}
    \label{fig:strain}
\end{figure}

SNR of the GW detector can be calculated by
\begin{equation}
\rho=\sqrt{\sum_{n=1}^{N_{\rm d}}(\rho_{n})^2},
\label{euqa:rho}
\end{equation}
with $\rho_{n}=\sqrt{( \tilde{h}_{n}|\tilde{h}_{n})}$.
$N_{\rm d}$ is the number of independent interferometers, with $N_{\rm d}$ = 3 for ET and $N_{\rm d}$ = 1 for CE1 and CE2.
The inner product is defined as
\begin{equation}
\left({a|b}\right)=4\int_{f_{\rm lower}}^{f_{\rm upper}}\frac{ a^\ast(f) b(f)}{S_{\rm n}(f)}{\rm d}f,
\label{euqa:product}
\end{equation}
where $f_{\rm lower}$ is the lower cut-off frequency ($f_{\rm lower}=1$ Hz for ET and $f_{\rm lower}=5$ Hz for two CEs), $f_{\rm upper}=2/(6^{3/2}2\pi M_{\rm obs})$ is the frequency at the last stable orbit
with $M_{\rm obs}=(m_1+m_2)(1+z)$ \cite{Zhao:2010sz}, and $S_{\rm n}(f)$ is the one-side noise power spectral density (PSD). 
We adopt PSDs of ET and CEs from Refs.~\cite{Hild:2010id,Evans:2021gyd}, shown in Fig.~\ref{fig:strain}. In our simulation, we assume that the mock GW signals with $\rho>8$ are detectable. By calculating SNRs of GW events in the GW catalog simulated in Sec.~\ref{sec:SIMULATION OF THE GW EVENT CATALOG}, we find that ET, CE1, and ET2CE can detect all GW events at $z \leq 0.3$. 

We use a $9\times 9$ FIM to estimate the measurement errors of the detected GW events' source parameters, including $d_{\rm L}$, $\mathcal{M}_{\rm c}$, $\eta$, $\theta$, $\phi$, $\iota$, $t_{\rm c}$, $\psi_{\rm c}$, and $\psi$. For a GW detector with $N_{\rm d}$ independent interferometers, FIM can be written as
\begin{equation}
{F}_{ij}=\sum_{n=1}^{N_{\rm d}} \left(\frac{\partial {\tilde{h}_{n}}}{\partial \theta_i}{\bigg |}  \frac{\partial {\tilde{h}_{n}}}{\partial \theta_j}\right),
\label{con:Fisher information matrix}
\end{equation}
where $\theta_i$ denotes the $i$-th parameter of the total nine source parameters. The covariance matrix of source parameters is approximately given by the inverse of FIM \cite{Vallisneri:2007ev},
\begin{align}
   {Cov}_{i j}=\left({F}^{-1}\right)_{i j}.
\label{con:covariance matrix}
\end{align}

\section{Search for host galaxies}\label{sec:Search}

To identify the host galaxies of GW events, we need to obtain the measurement errors of $d_{\rm L}$ and the angular locations of GW sources, and then construct a three-dimensional (3D) localization region.

We estimate the instrumental error of $d_{\rm L}$ via the FIM analysis, 
\begin{equation}
    \Delta d_{\mathrm{L}}^{\mathrm{inst}}=\sqrt{Cov_{d_{\rm L}d_{\rm L}}}.
\end{equation} 
We adopt a fitting formula to estimate the weak lensing error \cite{Tamanini:2016zlh},
\begin{equation}
    \begin{aligned}
        \Delta d_{\mathrm{L}}^{\mathrm{lens}}(z)=d_{\mathrm{L}}(z) \times 0.066\left[\frac{1-(1+z)^{-0.25}}{0.25}\right]^{1.8}.
    \end{aligned}
\end{equation}
The total error of $d_{\rm L}$ can be written as
\begin{align}
\Delta d_{\mathrm{L}}=\sqrt{(\Delta d_{\mathrm{L}}^{\mathrm{inst}})^{2}+(\Delta d_{\mathrm{L}}^{\mathrm{lens}})^{2}}.
\label{con:dl error}
\end{align}

The error of the solid angle, $\Delta \Omega$, is given by \cite{Zhao:2017cbb},
\begin{equation}
    \begin{aligned}
    \Delta \Omega=2\pi|\sin(\theta)|\sqrt{{Cov}_{\theta\theta}{Cov}_{\phi\phi}-{Cov}_{\theta\phi}^2}.
    \end{aligned}
\end{equation}

\begin{figure}
    \centering
    \includegraphics[width=8.6cm]{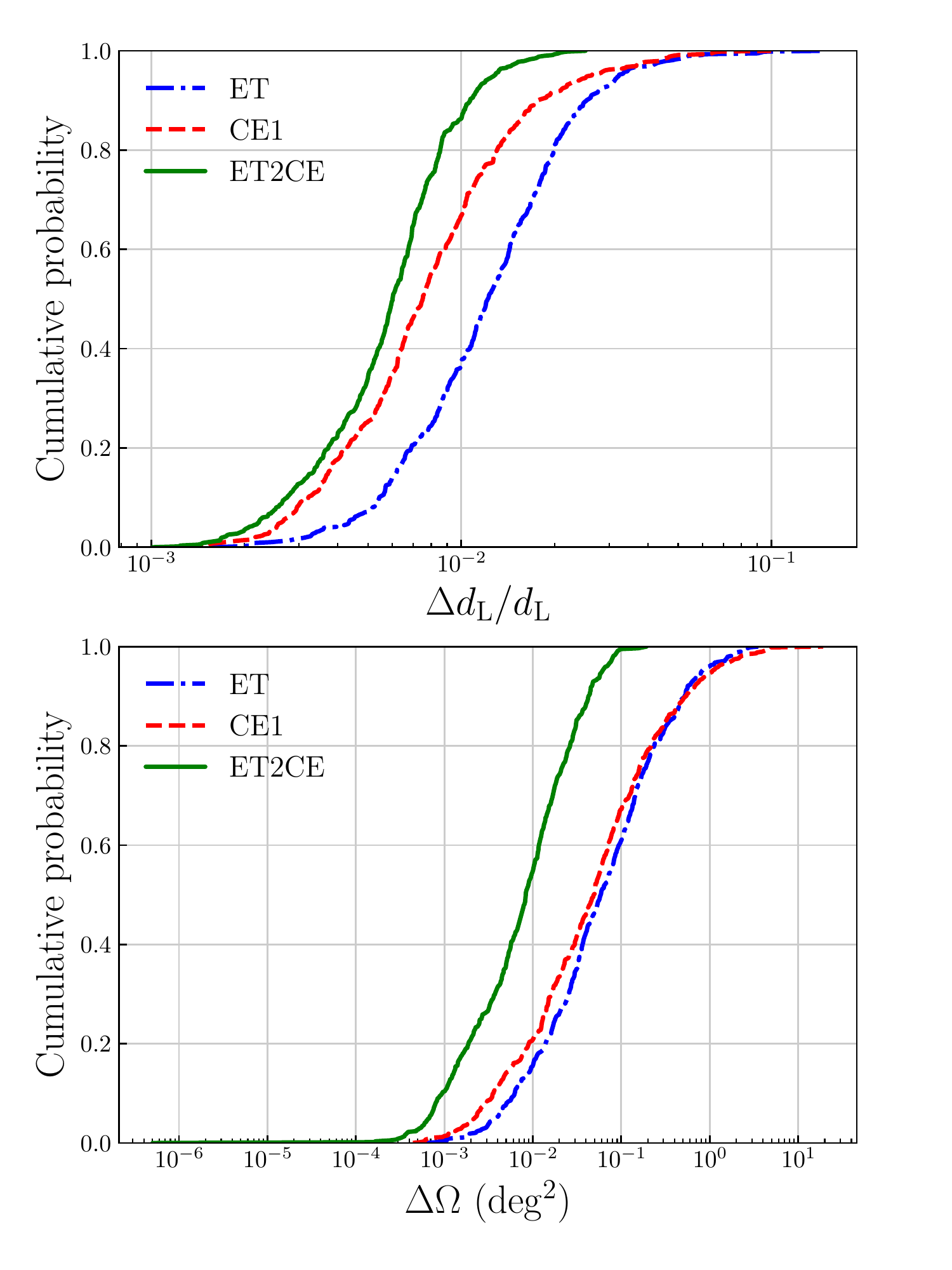}
    \caption{CDFs of $\Delta d_{\rm L}/d_{\rm L}$ and $\Delta \Omega$. The blue, red, and green lines represent the cases of ET, CE1, and ET2CE, respectively.}
    \label{fig:parameter_errors}
\end{figure}

In Fig.~\ref{fig:parameter_errors}, the upper and lower panels show the cumulative distribution function (CDF) of $\Delta d_{\rm L}/d_{\rm L}$ and $\Delta \Omega$, respectively. CDF of a variable $X$, evaluated at $x$, is the probability of $X\leq x$. 
The upper panel shows that, in terms of determining $d_{\rm L}$, ET has the worst capability among the considered GW detectors, with its CDF lower than CE1 and ET2CE.
The lower panel shows that, in terms of determining the angular location, ET and CE1 have similar capabilities, with their CDFs roughly overlapping; while ET2CE has the strongest capability, with a higher CDF distribution than both of them.

After obtaining $\Delta d_{\rm L}$ and $\Delta\Omega$, we can construct the 3D localization region. We define the 3D localization region of a GW event as the truncated cone with a radial range of [$\bar{d_{\rm L}}-3\Delta d_{\rm L}$, $\bar{d_{\rm L}}+3\Delta d_{\rm L}$] and an area of 3$\Delta \Omega$. $\bar{d_{\rm L}}$ represents the center value of the GW event's posterior distribution of the luminosity distance. Since the CSS-OS galaxy catalog contains the redshifts rather than the luminosity distances of galaxies, we need to convert the $d_{\rm L}$ range into the redshift range when matching galaxies to the GW source. We convert the lower and upper limits of the $d_{\rm L}$ range into the minimal redshift $z^{\rm min}$ and the maximal redshift $z^{\rm max}$ by setting $H_0$ to 60 and 80 $\ksm$, respectively. The values of 60 and 80 $\ksm$ correspond to the boundaries of the prior distribution of $H_0$ in our analysis, i.e., [60, 80] $\ksm$.

As for the range of the angular localization, following Ref.~\cite{Yu:2020vyy}, we use $\chi^2$ to describe it, given by
\begin{equation}
    \begin{aligned}
        \chi^2=(\theta-\bar{\theta},\phi-\bar{\phi})Cov'^{-1}\begin{pmatrix}\theta-\bar{\theta}\\ \phi-\bar{\phi}
\end{pmatrix},
    \end{aligned}
\end{equation}
where $Cov'$ is the 2-parameter covariance matrix including $\theta$ and $\phi$, obtained from the FIM analysis. $(\theta, \phi)$ and $(\bar{\theta}, \bar{\phi})$ correspond to the angular locations of an arbitrary galaxy and the true host galaxies of the GW event, respectively. $\chi^2$ quantifies the deviation of an arbitrary galaxy's angular location from the GW event's true angular location. The galaxies with $\chi^2 \leq 9.21$ (99\% confidence region) are roughly considered to be within the area of 3$\Delta\Omega$.

Finally, we can obtain the potential host galaxies of the GW event by selecting the galaxies in the redshift range $[z^{\rm min},z^{\rm max}]$ and with $\chi^2\leq9.21$. It is worth noting that there may be several GW events whose 3D localization regions are beyond the mock CSS-OS galaxy catalog, resulting in some potential host galaxies being neglected. To address this incompleteness, we only consider the GW events whose 3D localization regions are entirely within the redshift range of $z\leq0.3$ and the 17500 deg$^2$ coverage area of CSS-OS, and hence in our research, we consider 325, 334, and 341 GW events for ET, CE1, and ET2CE, respectively, in the 5-year observation.

\begin{figure}
    \centering
    \includegraphics[width=8.6cm]{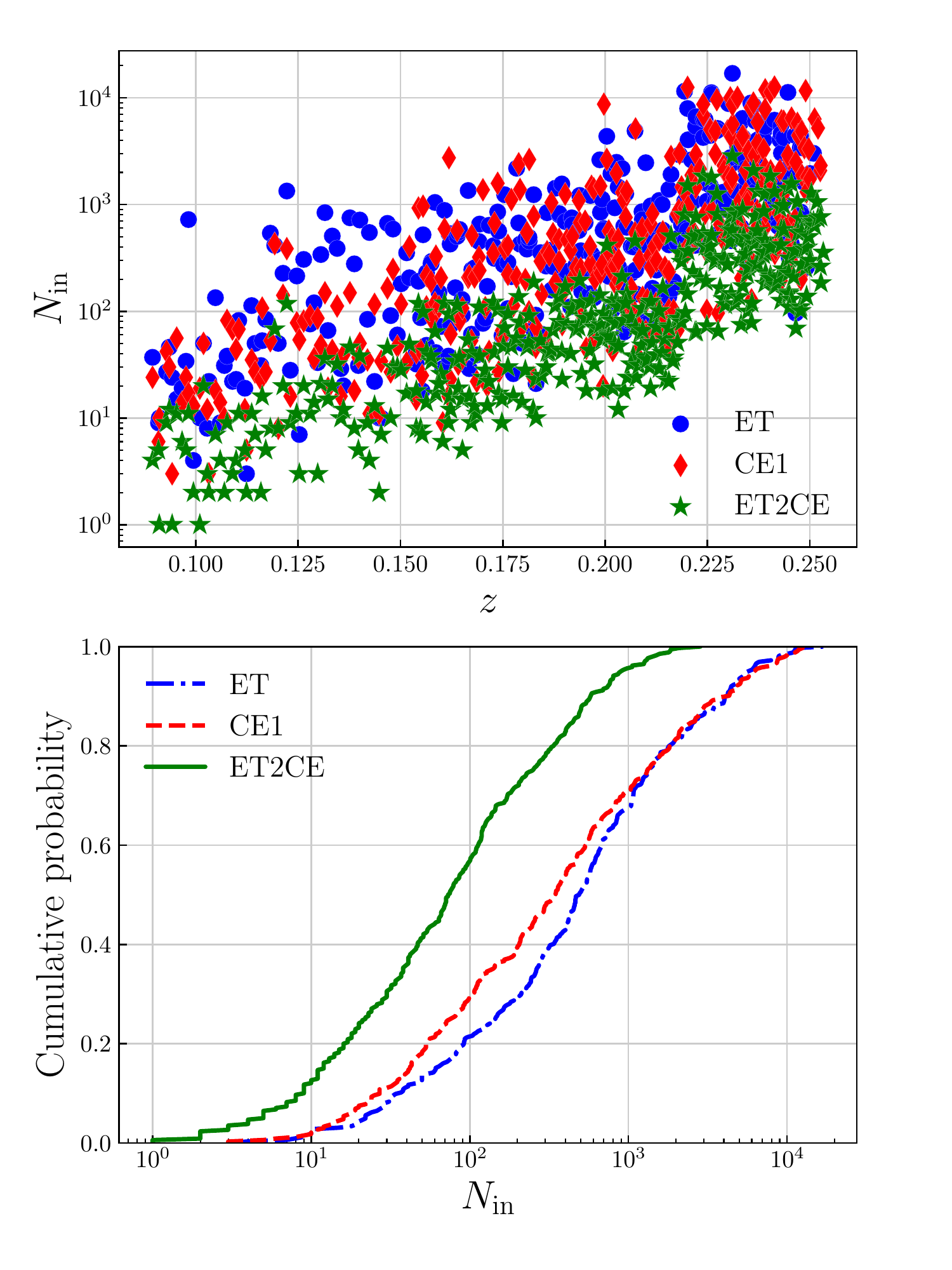}
    \caption{Distribution of $N_{\rm in}$ against $z$ (upper panel) and CDF of $N_{\rm in}$ (lower panel). The blue, red, and green colors represent the cases of ET, CE1, and ET2CE, respectively.}
    \label{fig:Nin_and_Nin}
\end{figure}

We define $N_{\rm in}$ as the number of potential host galaxies of the GW event. 
In the upper and lower panels of Fig.~\ref{fig:Nin_and_Nin}, we plot the scatter plot and the CDF plot of $N_{\rm in}$ for different GW detector cases, respectively. 
The upper panel shows that $N_{\rm in}$ increases with the redshift, indicating that the GW events at higher redshift have more fake potential host galaxies.
The lower panel shows that the CDF of ET2CE (the green solid line) is higher than that of ET and CE1, indicating that ET2CE can determine the host galaxies for the GW events more accurately.

\section{Constraints on the Hubble constant}\label{sec:baye}

We use the Bayesian method to infer $H_{0}$. The posterior probability of $H_{0}$ is given by
\begin{equation}
    \begin{aligned}
   p(H_0|\{d_{\rm GW}\},\{d_{\rm EM}\})\propto p(\{d_{\rm GW}\},\{d_{\rm EM}\}|H_0)p(H_0),
    \label{con:bayesian total}
    \end{aligned}
\end{equation}
where $\{d_{\rm GW}\}$ and $\{d_{\rm EM}\}$ represent the GW observation data set and the EM observation data set, respectively.
$p(\{d_{\rm GW}\},\{d_{\rm EM}\}|H_0)$ is the total likelihood function. $p(H_0)$ is the prior distribution of $H_0$ and we set it to be a uniform distribution between [60,~80] km s$^{-1}$ Mpc$^{-1}$. Assuming that the observations of the GW events are independent of each other, we express the total likelihood function as
\begin{equation}
    p(\{d_{\rm GW}\},\{d_{\rm EM}\}|H_0)=\prod_i^{N_{\rm GW}}p(d_{{\rm GW},i},d_{{\rm GW},i}|H_0),
\end{equation}
where $N_{\rm GW}$ is the number of the GW events.

For a single GW event, we can expand the likelihood as \cite{Chen:2017rfc,Mandel:2018mve}
\begin{equation}
    \begin{aligned}
        &p(d_{{\rm GW}},d_{{\rm GW}}|H_0)\\
        &=\iiiint \frac{1}{\beta(H_0)}p(d_{{\rm GW}}|d_{\rm L},\theta,\phi)p_{\rm pop}^{\rm GW}(d_{\rm L},\theta,\phi|H_0)\\
        & \times p(d_{{\rm EM}}|z,\theta,\phi)p_{\rm pop}^{\rm EM}(z,\theta,\phi|H_0){\rm d}d_{\rm L}{\rm d}\theta{\rm d}\phi{\rm d}z,
    \end{aligned}
\end{equation}
where $\beta(H_0)$ accounts for the selection effect and ensures that the likelihood integrates to unity.


Following Ref.~\cite{Yu:2020vyy}, we obtain the GW likelihood $p(d_{{\rm GW}}|d_{\rm L},\theta,\phi)$ based on the errors of the luminosity distance and the angular location discussed in Sec.~\ref{sec:Search},
\begin{equation}
    \begin{aligned}
        &p(d_{{\rm GW}}|d_{\rm L},\theta,\phi)=\frac{1}{\sqrt{2\pi}\sigma_{d_{\rm L}}}\exp\left[-\frac{1}{2}\left(\frac{d_{\rm L}-\bar{d_{\rm L}}}{\sigma_{d_{\rm L}}}\right)^2\right] \\
        &\times \frac{1}{2\pi |Cov'|}\exp\left[-\frac{1}{2}(\theta-\bar{\theta},\phi-\bar{\phi})Cov'^{-1}\begin{pmatrix}\theta-\bar{\theta}\\ \phi-\bar{\phi}
\end{pmatrix}\right].
    \end{aligned}
\end{equation}

We assume that the distribution of GW events $p_{\rm pop}^{\rm GW}(d_{\rm L},\theta,\phi|H_0)$ in the sky is isotropic and obtain the prior luminosity distance distribution $p_{\rm pop}^{\rm GW}(d_{\rm L}|H_0)$ by transforming the prior redshift distribution $p_{\rm pop}^{\rm GW}(z|H_0)$ of GW events via the $d_{\rm L}$-$z$ relation,
\begin{equation}
    \begin{aligned}
        p_{\rm pop}^{\rm GW}(d_{\rm L},\theta,\phi|H_0)&\propto p_{\rm pop}^{\rm GW}(d_{\rm L}|H_0)\\&\propto\delta[d_{\rm L}-d_{\rm L}(z,H_0)]p_{\rm pop}^{\rm GW}(z|H_0),
    \end{aligned}
\end{equation}
where $d_{\rm L}(z,H_0)$ is the theoretical luminosity distance calculated by Eq.~(\ref{con:distance-redshift relation}), and $p_{\rm pop}^{\rm GW}(z|H_0)$ is expressed as in Eq.~(\ref{con:pz}).

$p(d_{{\rm EM}}|z,\theta,\phi)$ is the EM likelihood in which we take into account the redshift error as a Gaussian form while ignoring the errors of the angular location,  
\begin{equation}
    \begin{aligned}
        p(d_{{\rm EM}}|z,\theta,\phi)=\sum_j^{N_{{\rm in}}}w_{j}N(\bar{z}_j,\sigma_{z,j})\delta(\theta-\bar{\theta}_j)\delta(\phi-\bar{\phi}_j),
    \end{aligned}
\end{equation}
where $w_i$ is the weight representing our prior knowledge of the probability of the $i$th potential host galaxy hosting the GW event. We assign equal weights to each galaxy, i.e., $w_i=1/N_{\rm in}$. $N(\bar{z}_j,\sigma_{z,j})$ is a Gaussian distribution centered at the measured $\bar{z}_j$. The form of $\sigma_{z}$ is discussed in Sec~\ref{subsec:redshift uncertainty}.

We assume that the prior distribution of galaxies $p_{\rm pop}^{\rm EM}(z,\theta,\phi|H_0)$ is uniform in the comoving volume,
\begin{equation}
    \begin{aligned}
        p_{\rm pop}^{\rm EM}(z,\theta,\phi|H_0)\propto p_{\rm pop}^{\rm EM}(z|H_0)\propto \frac{d_{\rm c}^2(z)}{H(z)},
    \end{aligned}
\end{equation}
where $p_{\rm pop}^{\rm EM}(z|H_0)$ is the prior redshift distribution of galaxies, and $d_{\rm c}(z)$ is the comoving distance.

In general, the GW and EM data are both affected by the selection effects, because they contain only the detected GW events and the observed galaxies. As discussed in Sec. \ref{subsec:completeness} and Sec. \ref{sec:FIM}, we set the GW detected threshold to $\rm{SNR}=8$ and only consider the dark sirens at $z<0.3$, ignoring the incompleteness of the CSS-OS photo-z catalog.
$\beta(H_0)$ is given by
\begin{equation}
    \begin{aligned}
        \beta(H_0)
        &=\iiiint p_{\rm det}^{\rm GW}(d_{\rm L},\theta,\phi)p_{\rm pop}^{\rm GW}(d_{\rm L},\theta,\phi|H_0)\\
        &\times p_{\rm det}^{\rm EM}(z,\theta,\phi)p_{\rm pop}^{\rm EM}(z,\theta,\phi|H_0){\rm d}d_{\rm L}{\rm d}\theta{\rm d}\phi{\rm d}z\\
        &= \iint p_{\rm det}^{\rm GW}(d_{\rm L})\delta[d_{\rm L}-d_{\rm L}(z,H_0)] p_{\rm pop}^{\rm GW}(z|H_0)\\
        &\times p_{\rm det}^{\rm EM}(z)p_{\rm pop}^{\rm EM}(z|H_0)
        {\rm d}z{\rm d}d_{\rm L}\\
        &= \int p_{\rm det}^{\rm GW}\left[d_{\rm L}(z,H_0)\right]p_{\rm pop}^{\rm GW}(z|H_0)\mathcal{H}(z_{\rm max}-z)\\
        &\times p_{\rm pop}^{\rm EM}(z|H_0){\rm d}z,
    \end{aligned}
\end{equation}
where $\mathcal{H}(z_{\rm max}-z)$ is the Heaviside step function, with $z_{\rm max}=0.3$. $p_{\rm det}^{\rm GW}\left[d_{\rm L}(z,H_0)\right]$ represents the GW event's detection probability at $d_{\rm L}(z,H_0)$. Following Ref.~\cite{Gray:2019ksv}, we calculate it by marginalizing the other source parameters except $d_{\rm L}$,
\begin{equation}
    \begin{aligned}
    p_{\rm det}^{\rm GW}(d_{\rm L})&=\int p_{\rm det}^{\rm GW}(d_{\rm L}|\{\theta\})p^{\rm GW}_{\rm pop}(\{\theta\}|d_{\rm L}){\rm d}\{\theta\},
    \end{aligned}
\end{equation}
where $\{\theta\}$ represents the GW event's source parameters except $d_{\rm L}$. $p_{\rm det}^{\rm GW}(d_{\rm L}|\{\theta\})$ represents the detection probability of the GW event at $d_{\rm L}$ and with $\{\theta\}$, and its value is 1 or 0, denoting whether the GW event exceeds the detection threshold of SNR.
$p^{\rm GW}_{\rm pop}(\{\theta\}|d_{\rm L})$ is the prior distribution of $\{\theta\}$ at $d_{\rm L}$, assumed to be a uniform distribution.

Here we use the Monte-Carlo integration, 
\begin{equation}
    \begin{aligned}
        p_{\rm det}^{\rm GW}(d_{\rm L})\approx \frac{1}{N_{\rm samp}}\sum_{i=1}^{N_{\rm samp}}p_{\rm det}^{\rm GW}(d_{\rm L}|{\{\theta\}}_i),
    \end{aligned}
\end{equation}
with
\begin{equation}
    \begin{aligned}
        p_{\rm det}^{\rm GW}(d_{\rm L}|{\{\theta\}}_i)=\left\{\begin{array}{ll}
1, & \text { if } \rho_{i}>8, \\
0, & \text { otherwise. }
\end{array}\right.
    \end{aligned}
\end{equation}
We randomly select the other source parameters while keeping $d_{\rm L}$ fixed to create the samples, and the number of the samples, $N_{\rm samp}$, is set to 50000.

\begin{figure*}
    \centering
    \includegraphics[width=17.2cm]{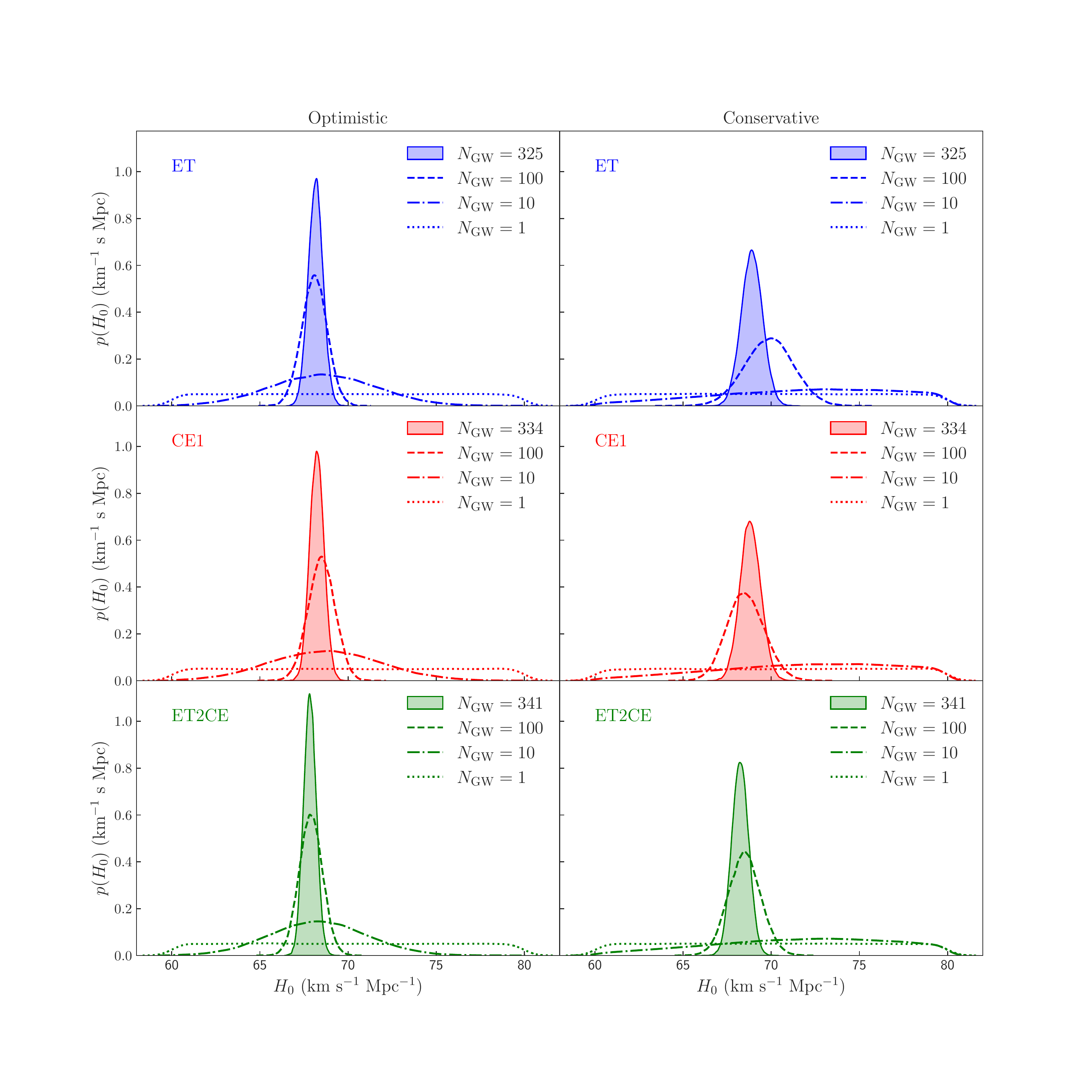}
    \caption{Posterior distributions of $H_0$ inferred from different numbers of GW events based on different GW detectors. The solid line with a shadow below represents the $H_0$ posterior distribution inferred from the combination of all GW events, and the dashed, dash-dot, and dotted lines represent $N_{\rm GW}=100$, $N_{\rm GW}=10$, and $N_{\rm GW}=1$, respectively. 
    Blue, red, and green colors represent ET, CE1, and ET2CE, respectively. The left and right columns correspond to the ``optimistic'' and ``conservative'' cases, respectively.
    }
    \label{fig:each_chain}
\end{figure*}


\begin{table*}
\centering
\caption{For each GW detector (column 1), we report the distribution of the number of potential host galaxies $N_{\rm in}$ and the constraint precision of $H_0$, denoted as $\Delta H_0/H_0$.
The second to fifth columns correspond to the fractions of GW events with different $N_{\rm in}$. The last column presents the 1$\sigma$ constraint precisions of $H_0$ for the ``optimistic'' case and the ``conservative'' case. The results of the ``conservative'' case are shown in parentheses.}\label{table:result}
\centering
\renewcommand{\arraystretch}{2}
\begin{tabular}{cccccc}
\hline\hline 
\makebox[0.15\textwidth][c]{Detector} & \makebox[0.15\textwidth][c]{$N_{\rm in}=1\ (\%)$} & \makebox[0.15\textwidth][c]{$N_{\rm in}\leq10\ (\%)$} & \makebox[0.15\textwidth][c]{$N_{\rm in}\leq100\ (\%)$} & \makebox[0.15\textwidth][c]{$N_{\rm in}\leq1000\ (\%)$} & \makebox[0.15\textwidth][c]{$\Delta H_0/H_0\ (\%)$} \\
\hline
ET & 0.0 & 3.08 & 21.54 & 67.38 & 0.62\ (0.89) \\
CE1 & 0.0 & 2.4 & 29.34 & 70.96 & 0.6\ (0.86) \\
ET2CE & 0.88 & 12.9 & 56.89 & 95.6 & 0.53\ (0.71) \\
\hline\hline
\end{tabular}

\end{table*}

\section{Results and discussions}
\label{sec:result}

In this section, we report the $H_0$ constraint results in different cases and make some relevant discussion. In Fig.~\ref{fig:each_chain}, we show the posterior distribution of $H_0$ inferred from different numbers of GW events based on ET, CE1, and ET2CE for the ``optimistic'' and ``conservative'' cases. In this figure, $N_{\rm GW}=325$, $N_{\rm GW}=334$, and $N_{\rm GW}=341$ represent the total numbers of GW events observed by ET, CE1, and ET2CE, respectively; $N_{\rm GW}=1$, $N_{\rm GW}=10$, and $N_{\rm GW}=100$ mean that we randomly select 1, 10, and 100 GW events from all GW events, respectively. We find that the posterior distribution of $H_0$ inferred from a single GW event (the dotted line) has extremely small peaks and looks almost flat, while the joint inference of multiple GW events can give a narrow and high $H_0$ posterior distribution, and the $H_0$ posterior distribution become narrower and higher as the number of GW events increases. We can also see that the ``optimistic'' case exhibits narrower and higher posterior distributions than the ``conservative'' case.

In Table~\ref{table:result}, we show the constraint precisions of $H_0$ inferred from all GW events based on ET, CE1, and ET2CE for the ``optimistic'' and ``conservative'' cases. The ``conservative'' case of ET can make $\Delta H_0/H_0$ reach $\sim$ 0.89\%. The constraint precisions of $H_0$ derived from the ``conservative'' case and the ``optimistic'' case of CE1 are 0.86\% and 0.60\%, respectively. Compared with ET, CE1 could improve the constraint precision of $H_0$ by about 3\%. This is mainly because the masses of BBHs we consider are roughly distributed between 10--45 $M_{\odot}$ and the GW frequencies in the inspire phases fall in $\mathcal{O} (10) \sim \mathcal{O} (10^2)$ Hz. In this frequency band, the sensitivity of CE1 is several times better than ET, which can be seen in Fig.~\ref{fig:strain}. These results are consistent with our predictions on the localization capabilities of the GW detectors in Sec.~\ref{sec:Search}.

When ET and two CEs form a detection network (ET2CE), this network makes the constraint precisions of $H_0$ improved by about 17.44\% (``conservative'' case) and 11.67\% (``optimistic'' case), compared with single CE1. These improvements are mainly due to the better localization capability of the GW detector network than a single detector. The errors of the luminosity distances and the solid angles of GW sources given by ET2CE are smaller than those given by CE1 by about 35.8\% and 92.3\%, and the smaller localization errors lead to smaller $N_{\rm in}$ of each GW event, providing a more accurate estimation for the GW event's redshift. In addition, as shown in Table~\ref{table:result}, CE1 and ET cannot uniquely identify the host galaxies ($N_{\rm in}=1$), while ET2CE allows us to uniquely identify the host galaxies for 0.88\% dark sirens, for which we can use them as bright sirens.

Compared with the state-of-the-art $H_0$ constraint results from the real observed dark sirens in GWTC-3 \cite{LIGOScientific:2021aug}, our $H_0$ constraint  results are better by about 95--97\%, mainly due to the following reasons. First, the CSS-OS photo-z catalog has nearly 100\% completeness up to $z\sim0.3$ and can reach as far as $z\sim4$, while {\tt GLADE+} is complete only up to $d_{\rm L}\sim47$ Mpc ($z\sim0.011$), and the completeness falls to 20\% at $d_{\rm L}\sim800$ Mpc ($z\sim0.167$). As is discussed in Ref.~\cite{Gray:2019ksv}, the better completeness performance leads to better $H_0$ constraint results with the same GW data. The larger survey depth of CSS-OS galaxy catalogs allows us to take more GW events into consideration. Second, the average galaxy redshift uncertainty in CSS-OS photo-z catalog is also smaller than that of {\tt GLADE+} by $\sim40\%$. Third, we consider 3G GW detectors, whose localization capabilities are better than the second-generation (2G) detectors. As predicted in Ref.~\cite{Yu:2020vyy}, compared with 2G GW detector networks, 3G GW detector networks could reduce the instrumental error of $d_{\rm L}$ by about four orders of magnitude.

To confirm our estimations, we compare our results with other papers that forecast the constraints on $H_0$ from dark sirens observed by the 3G GW detectors. Refs.~\cite{Yu:2020vyy,Borhanian:2020vyr} give $\Delta H_0/H_0<10^{-4}$ based on several years of observations by the 3G GW detector network. Our results are worse by 1--2 orders of magnitude compared with theirs. The main reason is that we utilize the latest BBH population distribution inferred from GWTC-3, finding fewer BBHs at $z\leq0.1$ than their assumption. We also consider the galaxy redshift uncertainty in the CSS-OS galaxy catalogs. In addition, the method of Ref.~\cite{Yu:2020vyy} is to search for host galaxy groups rather than host galaxies, which has more advantages in determining the redshifts of dark sirens and hence improves the constraints on $H_0$. Our findings generally agree with the results reported in Ref.~\cite{Zhu:2023jti}, which provides the cosmological forecasts for the network composted of ET and CE1 and focuses on the systematic error of $H_0$ arising from the incorrect spatial localization. Our results are the first prediction of the constraint on $H_0$ for the synergy between CSS-OS and 3G GW detectors.

In addition to the ground-based GW detectors, several papers studied the roles of the future space-borne GW observatories and the pulsar timing arrays (PTAs) in the dark-siren cosmology. Ref.~\cite{Zhu:2021aat} forecasted that TianQin could constrain $H_0$ to a precision of 4\% -- 7\%, and the TianQin-LISA network could make the precision achieve 1.7\%. Ref.~\cite{Wang:2020dkc} forecasted that the LISA-Taiji network can constrain $H_0$ to a 1\% precision. Ref.~\cite{Wang:2022oou} forecasted that using PTAs in the era of the Square Kilometre Array (SKA) may observe $\sim40$ dark sirens in 10 years and make the measurement precision of $H_0$ reach 1.8\%. We expect that CSS-OS will provide these space-borne GW detectors and PTAs with suitable galaxy catalogs, as it has large redshift coverage and high redshift measurement accuracy.

It is worth noting that we did not take into account the impacts of GW sources' peculiar velocities on the $d_{\rm L}$ measurements when we obtained the main results in this paper. To enhance the confidence of our results, we also make further calculations and obtain some additional results by taking into account the error on $d_{\rm L}$ introduced by GW sources' peculiar velocities. We add the peculiar-velocity error [$\Delta d_{\rm L}^{\rm pv}(z)$] into Eq.~(\ref{con:dl error}) and then it becomes $\Delta d_{\mathrm{L}}=\sqrt{(\Delta d_{\mathrm{L}}^{\mathrm{inst}})^{2}+(\Delta d_{\mathrm{L}}^{\mathrm{lens}})^{2}+(\Delta d_{\mathrm{L}}^{\mathrm{pv}})^{2}}$, with $\Delta d_{\rm L}^{\rm pv}(z)=d_{\rm L}(z) \times\left[1+\frac{c(1+z)^{2}}{H(z) d_{\rm L}(z)}\right] \frac{\sqrt{\left\langle v^{2}\right\rangle}}{c}$ \cite{Kocsis:2005vv}. Here the root mean square peculiar velocity is set to $\sqrt{\left\langle v^{2}\right\rangle}=500\ {\rm km\ s^{-1}}$ \cite{He:2019dhl}. We show these additional results in Table~\ref{table:result2} and find that including the peculiar-velocity error on $d_{\rm L}$ has a negligible effect on $N_{\rm in}$ and decreases the constraint precisions of $H_0$ by around 8\%--16\%. Even for the worst two cases (the ``conservative'' cases of ET and CE1), the precisions of $H_0$ could reach 1.01\% and 1.00\%, respectively. In the other cases, all the precisions of $H_0$ reach the sub-percent level, meeting the standard of precision cosmology.


\begin{table*}
\centering
\caption{Same as Talble~\ref{table:result} except considering the additional peculiar-velocity error in the $d_{\rm L}$ measurements.}\label{table:result2}
\centering
\renewcommand{\arraystretch}{2}
\begin{tabular}{cccccc}
\hline\hline 
\makebox[0.15\textwidth][c]{Detector} & \makebox[0.15\textwidth][c]{$N_{\rm in}=1\ (\%)$} & \makebox[0.15\textwidth][c]{$N_{\rm in}\leq10\ (\%)$} & \makebox[0.15\textwidth][c]{$N_{\rm in}\leq100\ (\%)$} & \makebox[0.15\textwidth][c]{$N_{\rm in}\leq1000\ (\%)$} & \makebox[0.15\textwidth][c]{$\Delta H_0/H_0\ (\%)$} \\
\hline
ET & 0.0 & 2.85 & 22.15 & 68.35 & 0.67\ (1.01) \\
CE1 & 0.0 & 2.14 & 29.36 & 72.17 & 0.65\ (1.00) \\
ET2CE & 0.91 & 13.29 & 58.61 & 96.07 & 0.59\ (0.86) \\
\hline\hline
\end{tabular}

\end{table*}

\section{Conclusion}

GW standard sirens are a late-universe cosmological probe with great potential to measure absolute cosmological distances and constrain cosmological parameters with the $d_{\rm L}$-$z$ relation. The dark siren method allows us to measure cosmological parameters using CBCs without EM counterparts. In this paper, we study the capabilities of the 3G GW detectors, together with the CSS-OS galaxy catalog, to measure the Hubble constant via the dark siren method.

First, we mock the CSS-OS galaxy catalog and estimate its completeness based on the Schechter function. Second, we simulate the GW events of the 5-year observation of ET, CE1, and the ET2CE network according to the population distribution of BBHs inferred from GWTC-3. Third, we estimate the errors of the source parameters with the FIM analysis and obtain the 3D localization regions. With the 3D localization region, we search for the potential host galaxies of the GW events in the mock CSS-OS galaxy catalog. Finally, we employ the Bayesian method to infer $H_0$.

Our results show that around 300 dark sirens from the 3G GW detectors and the CSS-OS galaxy catalog can constrain $H_0$ well. The constraint precisions of $H_0$ given by ET are 0.89\% (``conservative'' case) and 0.62\% (``optimistic'' case), and those given by CE1 are 0.86\% (``conservative'' case) and 0.60\% (``optimistic'' case). CE1 makes the precision of $H_0$ improved by about 3.37\% and 3.23\%, compared with ET. When ET and two CEs form a network, due to the much better localization capability, ET2CE makes the precisions of $H_0$ achieve 0.71\% (``conservative'' case) and 0.53\% (``optimistic'' case), improved by 17.44\% and 11.67\% compared with the results given by a single CE1. The constraint precisions of $H_0$ will decrease by about 8\%--16\% if we consider the peculiar-velocity error in the $d_{\rm L}$ measurements. 

We find that CSST could improve the quality of dark sirens in the following aspects. (i) The galaxy numbers of the CSST photometric and spectroscopic catalogs are about 100 and 10 times more than the current {\tt GLADE+} catalogs, respectively, helping to improve the completeness of galaxy catalogs and identify the true host galaxies of GW events. (ii) CSST could observe the galaxies at higher redshifts and make the completeness of the galaxy catalogs extend to $z\sim 0.3$, reducing the uncertainties arising from the galaxy incompleteness. (iii) The average galaxy redshift uncertainty in CSS-OS photo-z catalog is smaller than that of {\tt GLADE+} by $\sim40\%$, directly improving the measurement precision of $H_0$ via the $d_{\rm L}$-$z$ relation. We conclude that the synergy between CSST and future GW observations has great potential in precisely measuring the Hubble constant.

\label{sec:summary}

\begin{acknowledgments}
We are grateful to Yan Gong, Fu-Ren Deng, Mu-Xin Liu, Ji-Ming Yu, and Yue Shao for fruitful discussions.
This work was supported by the National SKA Program of China (Nos. 2022SKA0110200 and 2022SKA0110203), the National Natural Science Foundation of China (Nos. 11975072, 11875102, and 11835009), the science research grants from the China Manned Space Project (No. CMS-CSST-2021-B01) and the 111 Project (No. B16009).
\end{acknowledgments}

\bibliography{CSST}

\end{document}